# Thick GEM-like multipliers - a simple solution for large area UV-RICH detectors

R. Chechik*, A. Breskin and C. Shalem

*Dept. of Particle Physics, The Weizmann Institute of Scienc, 76100 Rehovot, Israele*

**Abstract**

We report on the properties of thick GEM-like (THGEM) electron multipliers made of 0.4 mm thick double-sided Cu-clad G-10 plates, perforated with a dense hexagonal array of 0.3 mm diameter drilled holes. Photon detectors comprising THGEMs coupled to semi-transparent CsI photocathodes or reflective ones deposited on the THGEM surface were studied with Ar/$CO_2$ (70:30), Ar/$CH_4$ (95:5), $CH_4$ and $CF_4$. Gains of ~$10^5$ or exceeding $10^6$ were reached with single- or double-THGEM, respectively; the signals have 5-10 ns rise times. The electric field configurations at the THGEM electrodes result in an efficient extraction of photoelectrons and their focusing into the holes; this occurs already at rather low gains, below 100. These detectors, with single-photon sensitivity and with expected sub-millimeter localization, can operate at MHz/mm$^2$ rates. We discuss their prospects for large-area UV-photon imaging for RICH.

## 1. Introduction

RICH detectors have become routine tools for relativistic-particle identification in large accelerator-based particle and heavy-ion experiments. The high counting rate and large multiplicities in some experiments impose harsh requirements on the design and operation of RICH system and on the photon detectors in particular. They are required to have fast response and excellent detection efficiency for single photons, and at the same time minimum sensitivity to ionizing background radiation. They are often required to have very large effective area, of several square meters, or to operate within non-vanishing magnetic fields. Gaseous detectors are good candidates for this assignment. Though wire chambers have been playing an important role in the detection of single UV photons, their sensitivity (gain) is limited by secondary photon- and ion-feedback effects. The latter also cause photocathode (PC) damage.

Good candidates for photon detectors in RICH are gaseous photomultipliers where photoelectrons (PE) are multiplied within holes. It has been demonstrated that such mode of operation strongly reduces secondary effects and results in high gain and fast signals. An array of independent holes provides true pixilated radiation localization. The idea has been studied for a large variety of applications throughout the last three decades [1]. The most extensively studied and applied hole-array avalanche multiplier is the Gas Electron Multiplier (GEM) [2], made of 50 μm Kapton® foil perforated with 50-70 μm etched holes at ~140 μm pitch. The GEM's most important properties are the high gain of single or cascaded elements in a variety of gases including noble gases, the fast (ns) signals and the efficient coupling to gaseous or solid radiation convertors, as reviewed in [3]. One of the most interesting applications is that of single-photon imaging detectors for RICH, in particular, that of a cascaded GEM with a reflective photocathode deposited on top of the first GEM in the cascade [4,5,6]. Other detectors were proposed, with coarser hole-array multipliers, like glass Channel Plates and the G-10 made Optimized GEM [7].

* Corresponding author. Tel.: +972-8-934-4966; fax: +972-8-934-2611; e-mail: Rachel.chechik@weizmann,.ac.il.

The Thick GEM-like electron multiplier (THGEM) discussed here [8,1] is produced by simple industrial processes; holes are drilled in a double-face Cu-clad printed circuit board and the hole's rim is etched to prevent discharges. The THGEM might be regarded as a scale-up of standard GEM. However, the electric fields, the avalanche multiplication parameters (e.g. Townsend coefficient) and the electron and ion transport parameters (e.g. diffusion) do not scale up with the substrate thickness. Therefore, the THGEM properties cannot be straightforwardly derived from those of GEM, and a comprehensive new study was required [1]. In this article we present a concise summary of our study that concerns the THGEM potential application as a large-area UV-photon imaging detector for RICH.

## 2. The THGEMs and the experimental procedures

THGEM electrodes, of 1-8 $cm^2$, were produced from G-10 plates using industrial PCB processing of precise drilling and etching. Some multipliers were made of Kevlar, as discussed below. In all cases the etched copper left a 0.1mm space at the hole rim; it was found important for preventing discharges and therefore for reaching considerably higher gains. Out of the large assortment of tested THGEMs [1] (0.4 - 3.2 mm thick, 0.3 - 1.5 mm hole diameters and 0.7 - 4.0 mm pitch, respectively), the most appropriate for atmospheric pressure operation were found to be those of 0.4 mm thickness and 0.3 mm hole diameter. Unless written otherwise, all the data presented here were obtained from such an electrode, with a pitch of 0.7 mm (Fig. 1).

The experimental setups and procedures used for measuring the effective gain are described in [8,1]. The detectors incorporate semitransparent or reflective CsI PCs coupled to single or double THGEMs. The effective gain is the product of the true gain in the holes and the electron transfer efficiency (ETE), namely the efficiency to focus the PEs into the holes. The effective gain is obtained by comparing the current from the last electrode in the THGEM cascade to the PE current emitted from the PC. Such current-mode measurements do not permit separating the true gain and ETE. But at high THGEM gains the ETE is 100% and the effective gain obtained from current-mode measurements is identical to the true gain; at very low THGEM voltages, below the multiplication threshold, the effective gain value is equal to the ETE value.

A true separation of ETE and gain is achieved with the pulse-counting method, described in details in [9,10]. It is based on recording single electron pulses, in which case electron transfer inefficiency is directly translated to counting rate deficiency. We use a relative measurement, comparing the counting rate in the examined system to that recorded in a reference system known to have 100% ETE. This is done, of course, under exactly the same experimental conditions, with identical PC, UV-light illumination, and total pulse-gain and electronics chain. The pulse-counting method was used to obtain the transfer efficiency of the THGEM with either semitransparent or reflective PCs, in various gases. More details are given elsewhere [1].

## 3. Results

### *3.1. Effective gain*

Fig. 2 depicts the effective gain measured with the THGEM of Fig. 1, in various gases. The data was obtained with a reflective CsI PC deposited on the top of the THGEM. The maximal gain varies between $10^4$ in pure $CF_4$ and $10^5$ in $Ar/CH_4$ (95:5). A higher total gain was obtained by operating two THGEMs in cascade; they can be mounted ~5 mm apart with a high transfer field, up to 3kV/cm, applied between them. Fig. 3 shows the effective gain in $Ar/CH_4$ (95:5) and $Ar/CO_2$ (70:30), measured with a semitransparent CsI PC and a cascade of two identical THGEMs (of Fig. 1), equally biased, mounted 5 mm apart with a transfer field of 3kV/cm between them; the gain of a single element is shown for comparison. The total gain of the cascade seems to be higher than the product of the two individual elements' gains. As confirmed by electric-field calculations, this is due to the very high transfer field, which penetrates inside the THGEM holes and modifies the multiplication factor.

*3.2. Electron transfer efficiency (ETE)*

The ETE, the efficiency to focus the electrons into the holes, plays an important role in any hole-based gas multipliers, including the THGEM; it affects the effective gain and the total gain in cascaded operation mode. In particular the ETE defines the detection efficiency of single PEs. It depends on the point of origin of the PEs and on the electric fields - $E_{drift}$ above the THGEM and $E_{TGEM}$ inside the holes. The general trends are the same as in standard GEMs [11,3]: with electrons originating from the drift space above the THGEM (e.g. with a semi-transparent PC) it is important to have a large $E_{TGEM}/E_{drift}$ ratio; with electrons originating from a reflective PC on the top surface of the THGEM it is optimal to use $E_{drift}=0$ and have high $E_{TGEM}$. However, due to the larger and denser holes in the THGEM discussed in this article as compared to standard GEM, the ETE is reaching 100% already at rather low gains. This is shown in Fig. 4 for a reflective PC on the THGEM of Fig 1. Similarly, with a semitransparent PC the ETE reaches 100% already at gains of 10-100, depending on the gas [1].

The dense holes array in this THGEM leaves only 55% of the surface for the reflective PC; a THGEM with hole distance of 1mm, leaving 77% of the surface for the reflective PC (standard GEM has ~80%) was recently tested. The effective gain and ETE results are almost identical.

*3.3. Pulse rise-time and counting-rate capability*

Avalanche development within 0.4mm long holes results in typical pulse rise-time of 8-10ns, as shown in Fig 5. The pulses are almost as fast as that of GEMs [12].

The counting rate capability of the THGEM was measured with single PEs from a reflective PC. No effective-gain drop was observed up to ~1MHz/mm$^2$ with gain 4x10$^4$ [1]. For comparison, in standard GEM, space charge effects with x-rays were not seen up to ~0.3 MHz/mm$^2$ at a similar gain [13].

*3.4. Ion backflow*

Using a single THGEM coated with a reflective PC and a 0 field in the gap above it, 98% of the ions created in the holes are collected at the PC. In this case, similar to GEM, it is possible to reduce the ion backflow only by cascading at least two THGEMs.
Using a single THGEM and a semitransparent PC placed a few mm above it, the ion backflow fraction collected at the PC increases almost linearly with the field applied between the two (Fig 6). With 1kV/cm (sufficient for full PE extraction from the PC) under the conditions of fig 6, the ion backflow fraction is 10%. Using this THGEM as a second multiplier in a cascade, this fraction represents the amount of ions flowing to the first THGEM, of which part is further trapped on the first THGEM's metallic bottom electrode.

*3.5. Operation in $CF_4$*

The THGEM operation in $CF_4$ provides high gain, good ETE and fast pulses. However, this operation is problematic due to the high voltages required in this gas, causing sporadic sparking. We found that the latter caused an irreparable damage to the electrode, which limited the maximum attainable voltage. Subsequent discharges further reduced the maximum attainable voltage. We suspected that the $CF_4$ sparks etch the glass-fiber matrix of the G-10 material, and therefore produced a THGEM electrode out of Kevlar-based printed board. However, though visual inspection of the electrode did not reveal any defects, it was not possible to raise the voltages to the values required by the $CF_4$ gas, for reasons not yet understood. Other materials are presently considered, such as Teflon–based printed boards.

## 4. Discussion and prospects.

The newly introduced THGEM opens new possibilities for conceiving large-area UV-photon detectors of moderate (sub-millimeter) localization resolution. These have potential applications in various fields, including large-area Cherenkov radiation imaging detectors. UV-photon detectors in RICH devices do not require high localization resolutions but rather good single-photon sensitivity, low sensitivity to ionizing background and in some cases high counting-rate capability. Though multi-GEM photomultipliers with reflective PCs [5,14,15] seem adequate for RICH, the new THGEMs discussed here may offer a simpler and more economic solution. The study presented here, and in a more detailed article [1], on THGEM photomultipliers with semitransparent and reflective CsI PCs, indicates that such detectors can operate at high gain in a variety of gases including $CF_4$, though the issue of permanent damage due to sparks in $CF_4$ requires further investigation. In both operation modes the PEs are efficiently detected; the signals are fast and the counting rate capability is in the $MHz/mm^2$ range. As with multi-GEM based photomultipliers [4], THGEMs with reflective PCs deposited on their top surface, employing a slightly reversed drift field, have potentially good photon detection efficiency (due to the thick PC and high surface field) but low sensitivity to ionizing background, as demonstrated in [12]. In this operation mode, PEs are efficiently extracted and focused into the THGEM holes, while particle-induced electrons are deflected in the opposite direction; this feature might be invaluable for applications in accelerator physics. On the other hand, THGEMs with semitransparent PC have an advantage of reduced ion backflow, improving photocathode longevity and stability.

## Acknowledgments

The work was supported in part by the Israel Science Foundation, project No. 151/1 and by the USA-Israel Bi-national Science Foundation, project No. 2002240. C.S. was supported in part by the Fund for Victims of Terror of the Jewish Agency for Israel. A.B. is the W.P. Reuther Professor of Research in the peaceful use of atomic energy.

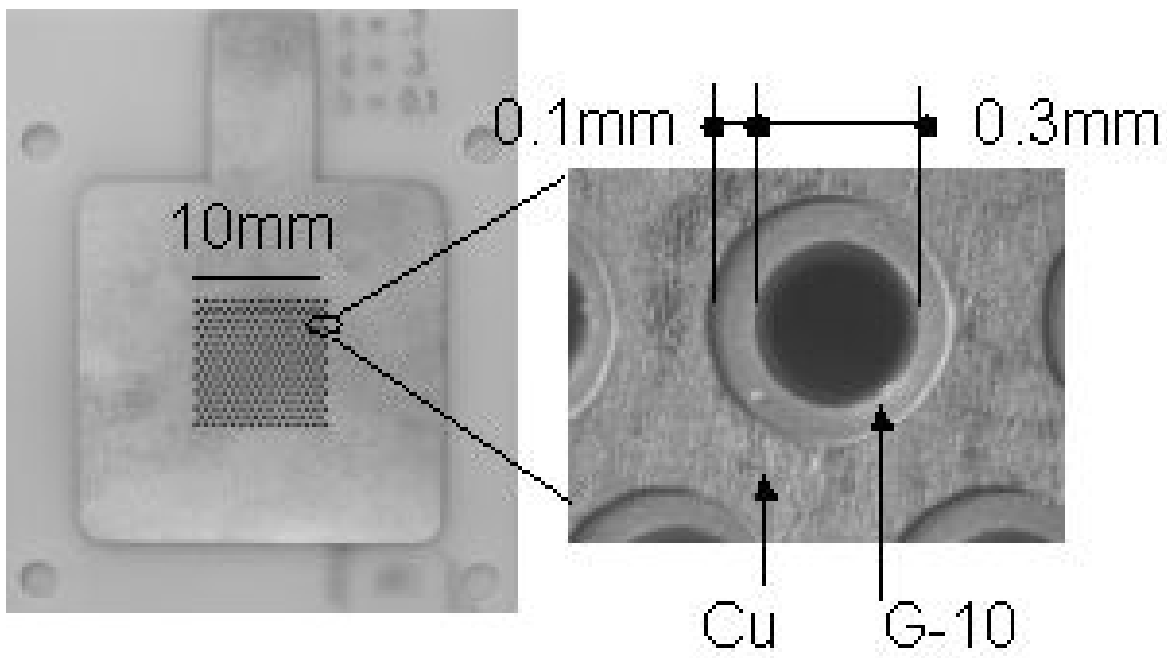

Figure 1. A photograph of a 0.4mm thick THGEM with 0.3mm holes and 0.7mm pitch. The enlarged part (right) shows the 0.1mm etched copper edge, preventing discharges at high potentials.

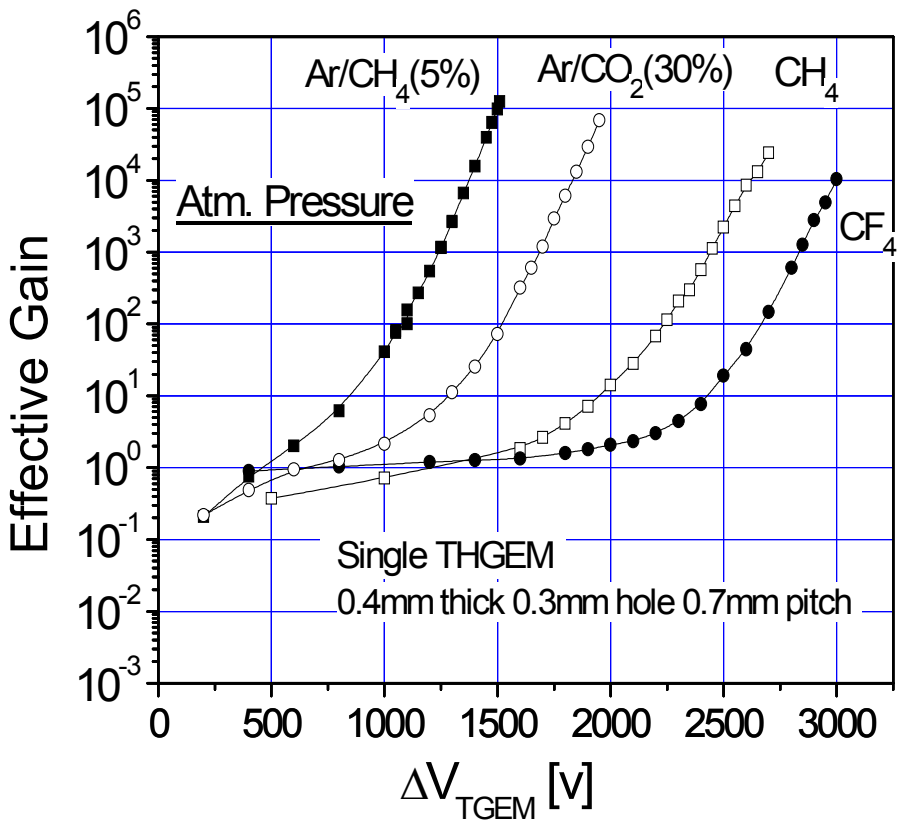

Figure 2. Effective gain of a single THGEM electrode with a reflective CsI photocathode, in various atmospheric pressure gases.

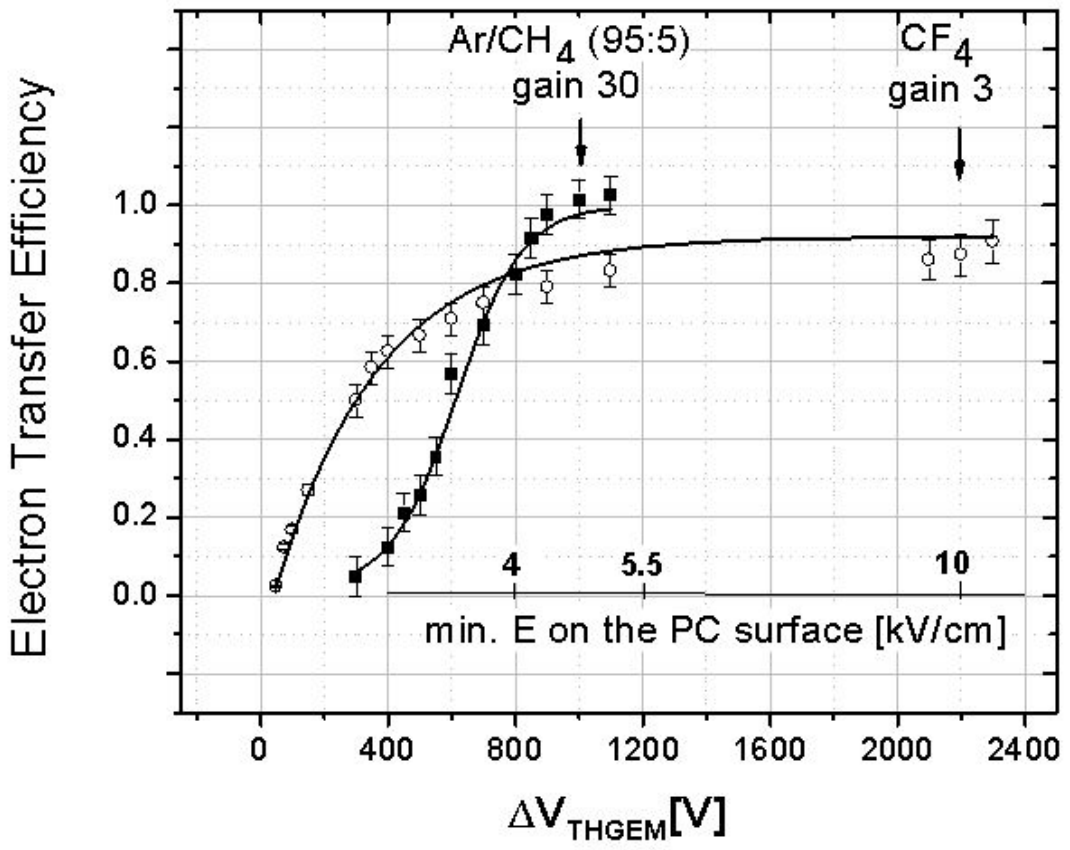

Figure 4. The electron transfer efficiency of a single THGEM coated with a reflective CsI photocathode. The corresponding THGEM gain and the minimal electric field on the CsI-coated surface (between holes) are indicated.

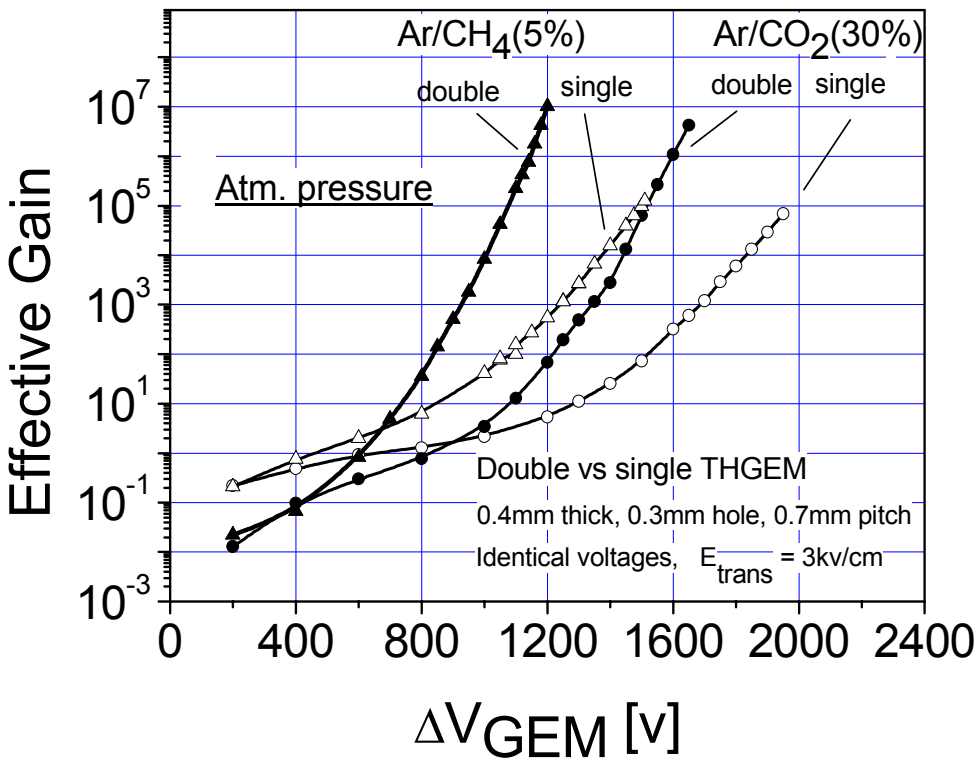

Figure 3. Effective gain of single- and double-THGEM detectors with identical electrodes equally biased, coupled to a semitransparent PC, in two Ar-based mixtures.

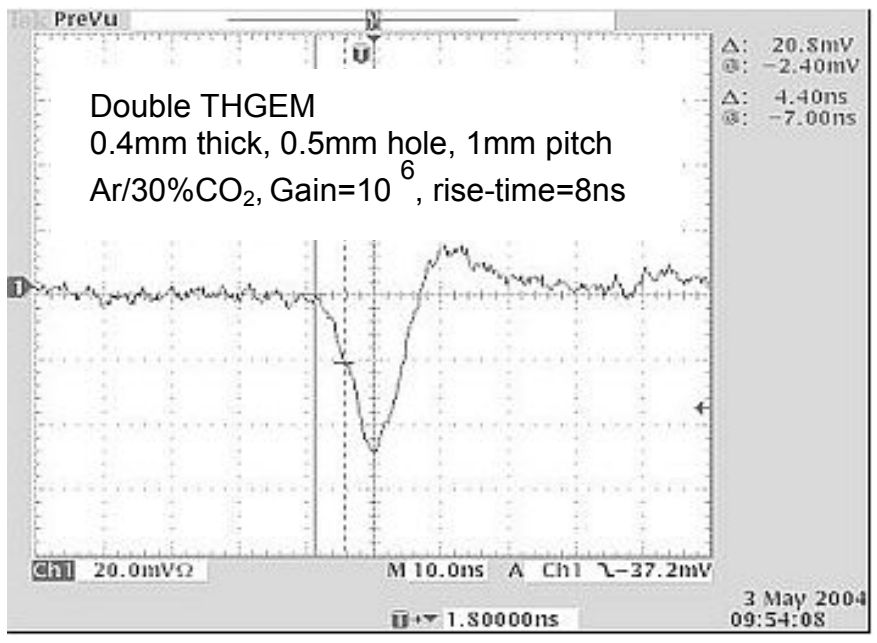

Figure 5.A single electron pulse captured with double THGEM in $Ar/CO_2$ (70:30), at a gain of $10^6$. The pulse has 8ns rise-time.

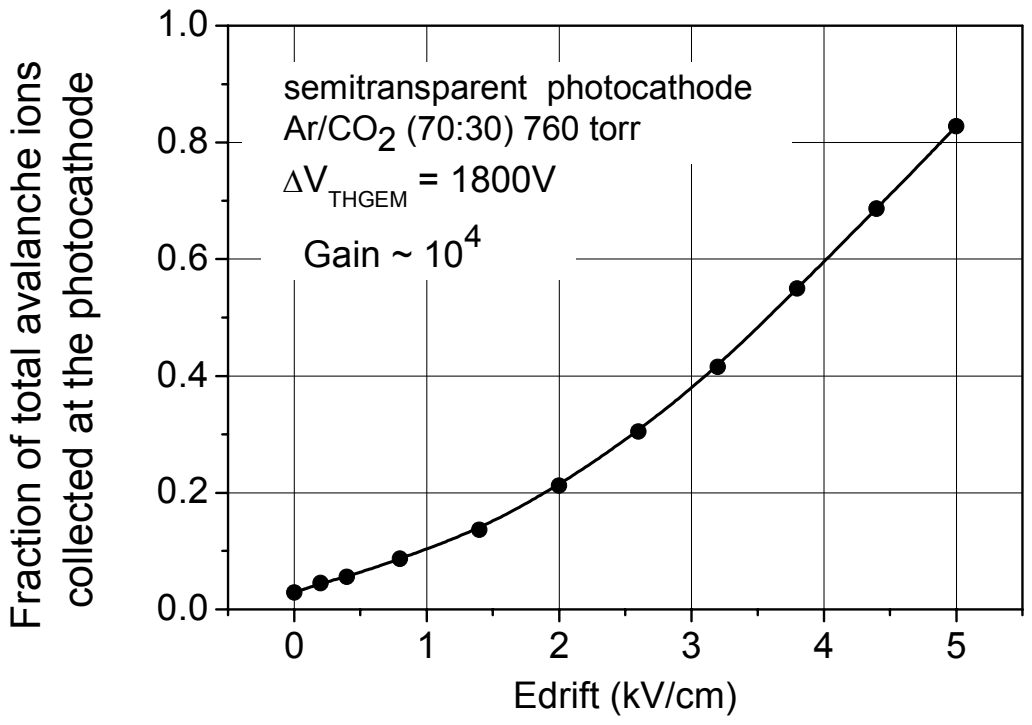

Figure 6. The fraction of total avalanche ions collected at a semitransparent photocathode coupled to a single THGEM. At field 0 it is 2%, and rises almost linearly with the drift field.